\documentclass[aps,prl,twocolumn,floats,floatfix]{revtex4}
\usepackage{ifthen}
\usepackage{ifpdf}
\usepackage{color}

\ifpdf
\usepackage{graphicx}
\usepackage{epstopdf}
\else
\usepackage{graphicx}
\usepackage{epsfig}
\fi

\usepackage{latexsym}
\usepackage{amsmath}
\usepackage{amssymb}
\usepackage{bm}
\usepackage{wasysym}

\newcommand{\ei}{\hat{a}}
\newcommand{\eidag}{\hat{a}^{\dag}}

\newcommand{\Lx}{\hat{L}_x}
\newcommand{\Ly}{\hat{L}_y}
\newcommand{\Lz}{\hat{L}_z}
\newcommand{\Lxy}{\hat{L}_{x,y}}

\newcommand{\hide}[1]{\textcolor{red}{[hidden]}}

\begin{document}

\title{Nonlinear phase-dynamics in a driven Bosonic Josephson junction}

\author{Erez Boukobza$^1$,  Michael G. Moore$^2$, Doron Cohen$^3$, and Amichay Vardi$^{1,4}$}
\affiliation{$^1$Department of Chemistry, Ben-Gurion University of the Negev, P.O.B. 653, Beer-Sheva 84105, Israel\\
$^2$Department of Physics \& Astronomy, Michigan State Univerity, East Lansing, Michigan 48824, USA\\
$^3$Department of Physics, Ben-Gurion University of the Negev, P.O.B. 653, Beer-Sheva 84105, Israel\\
$^4$ITAMP, Harvard-Smithsonian CFA, 60 Garden St., Cambridge, Massachusetts 02138, USA}

\begin{abstract}
We study the collective dynamics of a driven two mode Bose-Hubbard model in the Josephson interaction regime. The classical phase-space is mixed, with chaotic and regular components, that determine the dynamical nature of the fringe-visibility. For weak off-resonant drive, where the chaotic component is small, the many-body dynamics corresponds to that of a Kapitza pendulum, with the relative-phase $\varphi$ between the condensates playing the role of the pendulum angle. Using a master equation approach we show that the modulation of the inter-site potential barrier stabilizes the $\varphi=\pi$  'inverted pendulum' coherent state, and protects the fringe visibility.
\end{abstract}


\maketitle

%%%%%%%%%%%%%%%%%%%%%%%%%%%%%%%%%%%%%%%%%%%%%%%%%%%%%%%%%%%%%%%%%%%%%%%%%%%%%%%%%%%%%%%%%%%%%%
%%%%%%%%%%%%%%%%%%%%%%%%%%%%%%%%%%%%%%%%%%%%%%%%%%%%%%%%%%%%%%%%%%%%%%%%%%%%%%%%%%%%%%%%%%%%%%
%%%%%%%%%%%%%%%% INTRODUCTION %%%%%%%%%%%%%%%%%%%%%%%%%%%%%%%%%%%%%%%%%%%%%%%%%%%%%%%%%%%%%%%%

The Josephson effect \cite{Josephson62} is an unambiguous demonstration of macroscopic phase coherence between two coupled Bose ensembles. The experimental realization of  BEC Josephson junctions \cite{Cataliotti01,Anker05,Albiez05,Levy07} has led to observations of Josephson oscillations \cite{Javanainen86,Dalfovo96,Zapata98,Cataliotti01,Albiez05} as well as macroscopic self-trapping \cite{Smerzi97, Albiez05} and the equivalents of the ac and dc Josephson effect \cite{Giovanazzi00,Levy07}, present also in the superconductor \cite{Likharev79} or the superfluid Helium \cite{Pereverzev97,Sukhatme01} realizations. Beyond these mean-field effects, BEC Josephson systems allow for the observation of strong correlation phenomena, such as the collapse and revival of the relative phase between the two condensates \cite{Leggett98,Wright96,Javanainen97} which was observed with astounding precision in an optical lattice in Refs. \cite{Greiner02}, in a double-BEC system in Ref. \cite{Jo07}, and in a 1D spinor BEC in Ref. \cite{Widera08}.

The bosonic Josephson junction is often described by a two-mode Bose-Hubbard Hamiltonian (BHH)  \cite{BJM,Leggett01}, 
\begin{equation}
\label{Ham}
H=-K\Lx-{\cal E}\Lz+U\Lz^2~.
\end{equation}
Here $K$, {\cal E}, and $U$ are coupling, bias, and interaction energies. The SU(2) generarators $\Lx=(\eidag_1 \ei_2+\eidag_2\ei_1)/2$,  $\Ly=(\eidag_1\ei_2-\eidag_2\ei_1)/(2i)$, and $\Lz=(n_1 - n_2)/2$, are defined in terms of the boson on-site annihilation and creation operators $\ei_i$, $\eidag_i$, with the conserved total particle number $n_1+n_2=N\equiv 2\ell$. Three distinct interaction regimes are obtained, depending on the characteristic strength of interaction $u=UN/K$ \cite{BJM,Boukobza09}. The quasi-linear Rabi regime $|u|<1$, the strong-coupling Josephson regime $1<|u|<N^2$, 
and the extremely quantum Fock regime $u>N^2$. 

With adjustable parameters, the BHH can also realize an atom interferometer, in which the bias, ${\cal E}$, generates a phase-shift that can be measured by atom-number counting. The interaction term allows for the creation of non-classical input states, but also generates undesired phase-diffusion noise. Atom interferometers based on this model are of great current interest because they can potentially resolve phase-shifts below the standard quantum limit (SQL) of $\Delta\phi\ge1/\sqrt{N}$, and are limited instead by the Heisenberg fundamental limit $\Delta\phi\ge 1/N$. In such a device, a highly correlated initial state would be prepared in the Josephson or Fock regime, but the measurement would ideally be made in the Rabi regime. While tunable, e.g.  via Feshbach resonance, the interaction parameter $u$ will never be exactly zero. Thus understanding, and potentially harnessing, the dynamical effects of the interaction-induced nonlinearity will play a crucial role in designing such devices.

To model the nonlinear dynamics in the Josephson and Rabi regimes, it is convenient to define the action-angle variables $\ei=e^{i\varphi_i}\sqrt{n_i}$,  $\eidag=\sqrt{n_i}e^{-i\varphi_i}$. Using these definitions, the Hamiltonian (\ref{Ham}) can be rewritten in terms of the relative pair-number $n=(n_1-n_2)/2$ and relative-phase $e^{i\varphi}\equiv e^{i\varphi_1}e^{-i\varphi_2} $ operators,
\begin{eqnarray}
\nonumber
%\label{nphiHam}
H=Un^2-\mathcal{E}n-\frac{K}{2}
\left[\left(\frac{N}{2}{-}n\right)\left(\frac{N}{2}{+}n{+}1\right)\right]^{1/2} 
e^{i\varphi}+H.c.
\end{eqnarray}
In the Josephson interaction regime, states initiated with $n\ll N/2$ 
remain confined to this small population imbalance region, so that it is possible to use Josephson's approximated Hamiltonian   
\begin{equation}
\label{JosephsonHam}
H_{Josephson} = E_c (n-n_{\cal E})^2 - E_J \cos(\varphi)~,
\end{equation}
with $E_C=U$, $E_J=KN/2$, and $n_{\cal E}={\cal E}/(2U)$. 
This Hamiltonian matches that of a pendulum with `mass' ${M=1/(2U)}$ and frequency $\omega_J\approx\sqrt{KUN}$.   
The role of pendulum angle is played by the relative phase $\varphi$ between the two condensate modes, while the relative population imbalance, $n$, plays the role of angular momentum. One consequence of this analogy is that a relative-phase of $\varphi=0$ is classically stable (ground state of the pendulum), whereas a $\varphi=\pi$ relative-phase (inverted pendulum) is classically unstable. 

The term ``phase-diffusion" then describes the nonlinear effects generated by the anharmonic part of the $\cos(\varphi)$ potential. The many-body manifestation of this anharmonicity is the rapid loss of single-particle coherence for a coherent state prepared with a $\pi$ relative-phase, as opposed to the slow phase-diffusion of phase-locked condensates with $\varphi=0$ \cite{Boukobza09,Vardi}. Several recent works propose to control such phase-diffusion processes by means of external noise \cite{Khodorkovsky08,Witthaut08} or modulation of the Hamiltonian parameters to induce $\pi$ flips of the relative phase \cite{Bargill09}.

In this work, we build further on the pendulum analogy 
to explore the effect of oscillatory driving 
on the collective phase dynamics of the BHH. 
We consider two possible time-dependent driving fields,  ``vertical'' ($v$) and ``horizontal'' ($h$), given by
\begin{equation}
\label{vdrive}
V_{v,h}(t)=f(t){\hat W} = D_{v,h} \sin(\omega t+\phi) \ \Lxy~.
\end{equation}
Here `vertical' and `horizontal' are in reference to the pendulum model. The classical phase-pendulum motion is in the $L_x L_y$ equatorial plane of the BHH, with the $\varphi=0,\pi$ stationary points lying on the $L_x$ axis, making it the 'gravitation' direction of the pendulum. Hence, $V_v$  is equivalent to a vertical drive of the pendulum axis and $V_h$ corresponds to a horizontal drive. With respect to the two-mode BHH, the first type of driving $V_v(t)$ is a modulation of the hopping frequency $K$, 
which may be attained by changing the Barrier height, as illustrated in Fig.~1(a);
whereas $V_h(t)$ may be induced by means of shaking the double-well 
confining potential laterally, as illustrated in Fig.~1(b), thus effectively introducing 
the equivalent of an electromotive force in the oscillating frame.
It is customary to define the dimensionless frequency $\Omega\equiv\omega/\omega_J$,
and the dimensionless driving strength $q\equiv \sqrt{UN/K}(D/\omega)=D/(K\Omega)$. Fast and slow driving correspond to $\Omega\gg1$ and $\Omega\ll1$, respectively, whereas $q\gg\Omega$ and $q\ll \Omega$ correspond to strong and weak driving.

Within the 1D pendulum approximation, the angle variable $\varphi$ and the momentum $n$ 
are canonical conjugate variables.  It is well known \cite{Kapitza} that off-resonant, fast driving is effectively equivalent to the additional static `pseudo-potential ', $V^{\mbox{\tiny eff}}_{v,h}=\pm (1/4)q^2 K\ell\sin^2(\varphi)$, as illustrated in Fig.~1.  It is possible to further refine this effective description, adding momentum dependent terms, as described in \cite{Rahav03}.  For sufficiently strong (${q^2 > 2}$) vertical drive, the effective term $V^{\mbox{\tiny eff}}_{v}$ can stabilize the $\varphi=\pi$ inverted pendulum (Fig.~1(c)), an effect known as the Kapitza pendulum \cite{Kapitza}. By contrast, the effective term $V^{\mbox{\tiny eff}}_{h}$ can destabilize under the same conditions the $\varphi=0$ pendulum-down ground state,  
and generate two new degenerate quasi-stationary states (Fig.~1(d)). 
%%%%%%%%%%%%%%%%%%%%%%%%%%%%%%%%%%%%%%%%%%%%%%%%%%%%%%%%%%%%%%%%%%%%%%%%%%%%%%
%%%%%%%%%%%%%%%%%%%%%%%%%%%%%%%%%%%%%%%%%%%%%%%%%%%%%%%%%%%%%%%%%%%%%%%%%%%%%%
% chaos 

Generally, the driven BHH has a mixed classical phase-space structure similar to that of a kicked top  \cite{Haake87}, with chaotic regimes bound by KAM trajectories, making the bosonic Josephson junction a good system for studying quantum chaos \cite{Ghose01,Weiss08} along with the kicked-rotor realization by cold atoms in periodic optical lattice potentials \cite{KickedRotor}, ultracold atoms in atom-optics billiards \cite{Billiards}, and the recent realization of a quantum kicked top by the total spin of single $^{133}$Cs atoms \cite{Chaundry09}. The unique features of the driven BHH in this respect are: 
{\bf (i)} It offers a novel avenue of 'interaction-induced' chaos, 
which should be distinguished from the single-particle 'potential-induced' chaos 
that had been highlighted in past experiments with cold atoms; 
{\bf (ii)} The pertinent dynamical variables are relative-phase and relative-number, 
leading to nonlinear and possibly chaotic {\it phase} dynamics, which may be monitored 
via fringe-visibility measurements in interference experiments.

% fig1
%%%%%%%%%%%%%%%%%%%%%%%
\begin{figure}[t]
\centering
\includegraphics[angle=0,width=0.8\hsize]{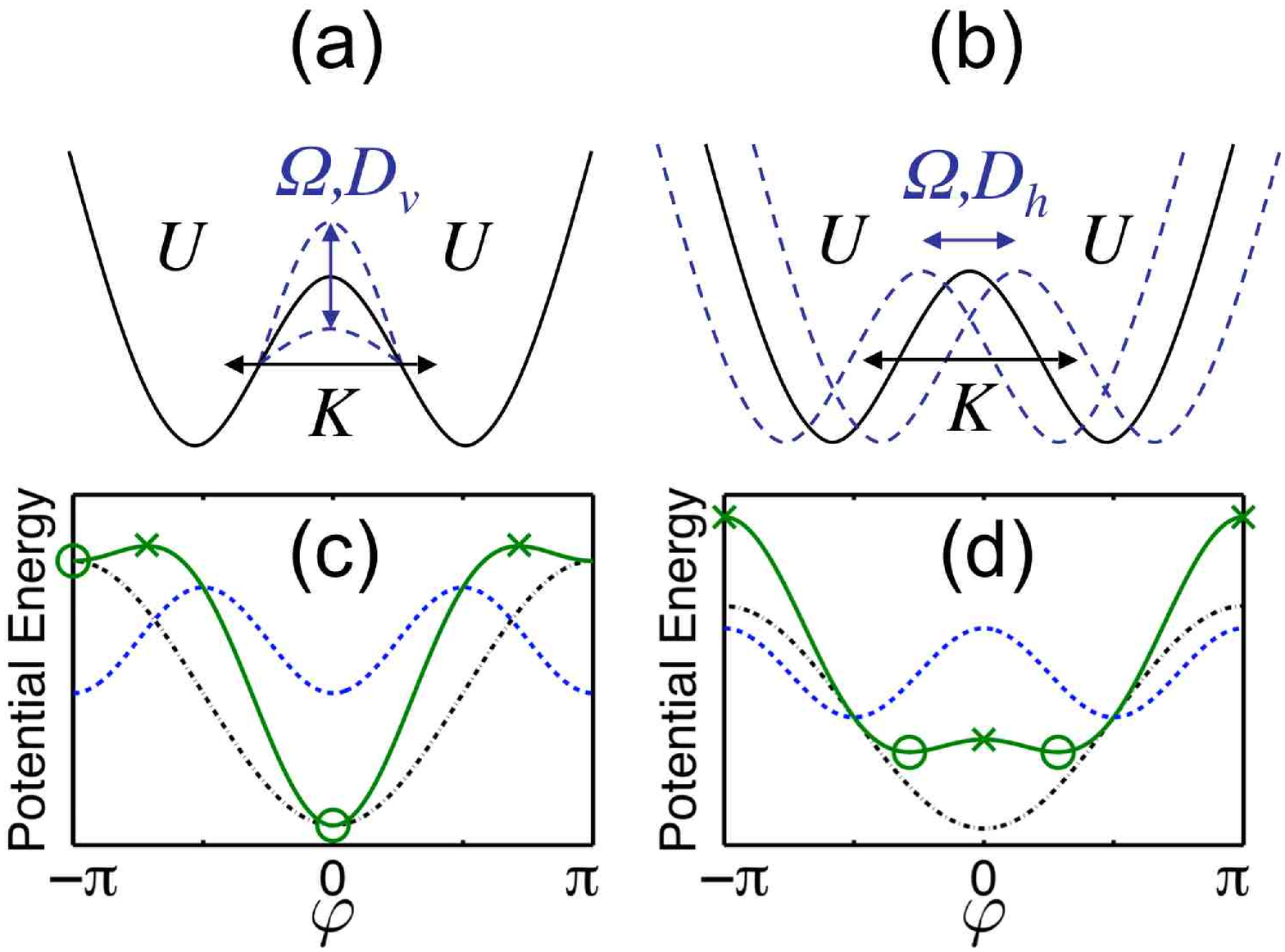}
\caption{(Color online) 
Schematics  of a driven Bose-Josephson junction: 
(a) The 'vertical' driving obtained by time-dependent modulation 
of the barrier height between the wells; 
(b) The 'horizontal' driving is via lateral shaking of the double-well potential; 
(c) The potential term in the Josephson Hamiltonian   
without driving (dash-dot), and with vertical driving (solid)  
which includes the effective potential (dashed).
Circles denote stable stationary points whereas 'x' denotes instabilities; 
(d) The same for horizontal driving. 
It should be noted that 'vertical'  and 'horizontal' refer to the motion of 
the pendulum axis in the Kapitza analogy, which incidentally match the direction of potential modulation}
\label{Schematics}
\end{figure}
%%%%%%%%%%%%%%%%%%%%%%

%fig2
%%%%%%%%%%%%%%%%%%%%%%%
\begin{figure}[t]
\centering
\includegraphics[angle=0,width=0.50\textwidth]{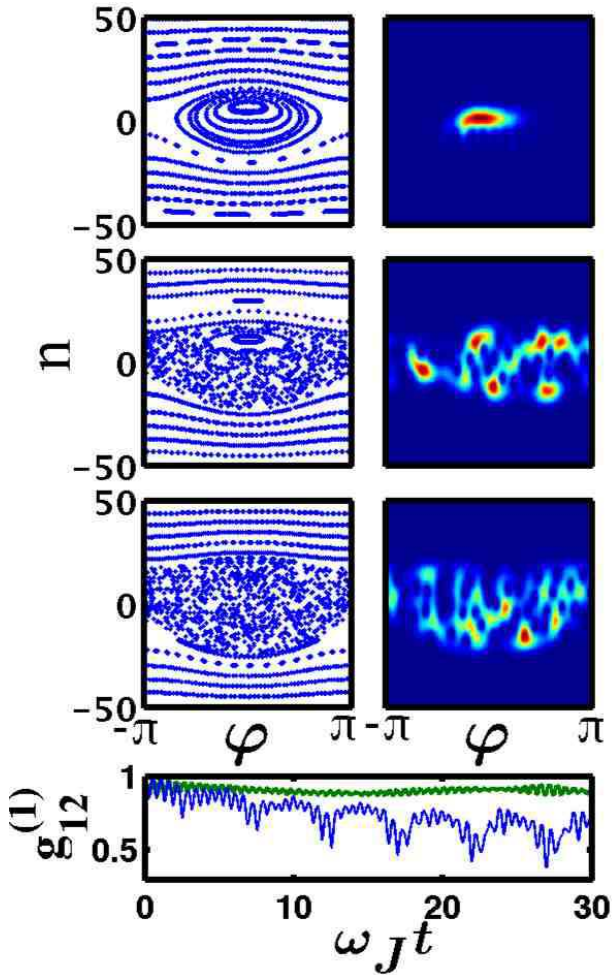}
\caption{(Color online)
Mean-field stroboscopic phase-space plots of {\em classical} trajectories (left), 
and a representative $N{=}100$ many-body {\em quantum} Hussimi distribution (right), 
for the dynamics that is generated by the BHH with $V_h(t)$ driving, 
assuming ${|0,0\rangle}$ coherent preparation.
The parameters are $u{=}30$, and $\Omega{=}1$. 
The strength of the driving is $q{=}0.1$ (top), and $q{=}0.5$ (middle), and $q{=}1.0$ (bottom). 
The evolution of the fringe-visibility is plotted in the lower panel  
for $q{=}0.1$ (bold solid, green) and $q{=}1.0$ (solid, blue). 
The circle indicates the time of the Hussimi plot.}
\label{HorizontalChaos}
\end{figure}
%%%%%%%%%%%%%%%%%%%%%%

Figure~2 shows representative results for near-resonant ($\Omega\approx 1$) horizontal driving. Stroboscopic Poincare plots of the classical (mean-field) evolution at drive-period intervals are shown on the left for varying drive strength, $q$, demonstrating the growth of the stochastic component to form a chaotic 'sea' surrounding regular 'islands' of non-chaotic motion.  
On the right, we represent the full many-body BHH evolution via the Husimi $Q$ function $Q(n,\varphi)=|\langle n,\varphi | \psi(t)\rangle|^2$, 
which provides visualization for the expansion of the time-dependent many-particle state $|\psi(t)\rangle$ 
in the spin coherent states $|n,\varphi\rangle$ basis.
%
%$|n,\varphi\rangle=\left[\sqrt{N/2+n}\eidag_1+e^{i\varphi}\sqrt{N/2-n}\eidag_2\right]^N|vac\rangle$
%
For weak driving the initial preparation $|0,0\rangle$ lies within a regular region of phase-space and retains its coherence. In contrast, for larger values of~$q$ the initial coherent state spreads quickly throughout the chaotic sea, resulting in a highly correlated many-body state, as manifest in the dynamics 
of fringe visibility $g_{12}^{(1)}(t)=\sqrt{\langle\Lx\rangle^2+\langle\Ly\rangle^2}/\ell$. 
Similar results, with a slightly different classical phase-space structure are obtained for vertical driving.

%%%%%%%%%%%%%%%%%%%%%%%%%%%%%%%%%%%%%%%%%%%%%%%%%%%%%%%%%%%%%%%%%%%%%%%%%%%%%%
%%%%%%%%%%%%%%%%%%%%%%%%%%%%%%%%%%%%%%%%%%%%%%%%%%%%%%%%%%%%%%%%%%%%%%%%%%%%%%
% kapitza

For an off-resonant weak driving with $\Omega\gg1$ and $q \ll \Omega$, 
the chaotic component of phase-space becomes hard to resolve.
In this case we obtain the many-body manifestation of the Kapitza pendulum physics \cite{Kapitza}, where a periodic vertical drive of the pendulum axis stabilizes the $\varphi=\pi$ inverted pendulum and a horizontal drive destabilizes its ground $\varphi=0$ state.

In generalizing the Kapitza pendulum physics into the context of the driven bosonic Josephson junction, we have to deal with two modifications: {\bf (i)} The phase-space of the full BHH is spherical and not canonical, as opposed to the truncated, cylindrical Josephson phase-space; {\bf (ii)} Realistically $f(t)$ may include a noisy component whose effect on the effective potential should be determined. 
We therefore re-derive the Kapitza physics for the full BHH, using a master equation approach rather than by the standard timescale separation methodology. The quantum state of the system is represented by the probability matrix $\rho$, satisfying ${d\rho/dt = i[\mathcal{H},\rho]}$ where ${ \mathcal{H}=H{+}f(t)W}$ with ${f(t)=\sin(\omega t+\phi)}$.  The small parameter is the driving period $\delta t=2\pi/\omega$ for harmonic drive, or the correlation time if $f(t)$ is noisy. Using a standard iterative procedure the difference 
${\rho(t+ \delta t)-\rho(t)}$ can be expressed to $1st$ order as an integral over ${[H{+}f(t')W,\rho]}$. 
The $2nd$ order adds a double integral over ${[H{+}f(t')W,[H{+}f(t'')W,\rho]]}$, and the $3rd$ order adds a triple integral over ${[H{+}f(t')W, [H{+}f(t'')W,[H{+}f(t''')W,\rho]]]}$. If $f(t)$ contains a noisy component, 
as in the standard master equation treatment, we obtain after integration over the second order contribution a diffusion term  $[W,[W,\rho]]$. In the familiar classical Focker-Planck context, with ${W=x}$, 
this terms takes the form ${\partial^2\rho(x,p)/\partial p^2}$. However, for a strictly periodic {\em noiseless} driving the 
time integration over a period vanishes, and evaluation to $3rd$ order is required. Integrating the $3rd$ order contribution over a period we get terms that can be packed as $[[W,[W,H]],\rho]$. Hence, the effective static potential is, 
\begin{equation}
\label{eff} 
V^{\mbox{\tiny eff}} \ = -\frac{1}{4\omega^2} \ [W,[W,H]]   
\end{equation}
Other terms also exist (producing the tilt of the islands in Figs.~3c,~4c), 
however they depend on the driving phase $\phi$, and vanish if the stroboscopic sampling is averaged.  
In the standard canonical case with ${\hat{W}=W(\hat{x})}$ 
this expression gives the familiar Kapitza result $[W'(x)]^2/(4M\omega^2)$ as in \cite{Rahav03}. 
For the BHH non-canonical spherical phase-space, with $W\propto L_x$ or $W \propto L_y$,  
it is straightforward to verify that Eq.(\ref{eff}) generates the expected 
Kapitza terms $L_y^2$ or $L_x^2$, as well as additional terms 
that slightly re-normalize the bare values of $U$ and $K$.

%%%%%%%%%%%%%%%%%%%%%%%%%%%%%%%%%%%%%%%%%%%%%%%%%%%%%%%%%%%%%%%%%%%%%%%%%%%%%%
%%%%%%%%%%%%%%%%%%%%%%%%%%%%%%%%%%%%%%%%%%%%%%%%%%%%%%%%%%%%%%%%%%%%%%%%%%%%%%
% numerics 

%fig3
%%%%%%%%%%%%%%%%%%%%%%%
\begin{figure}[t]
\centering
\includegraphics[angle=0,width=0.50\textwidth]{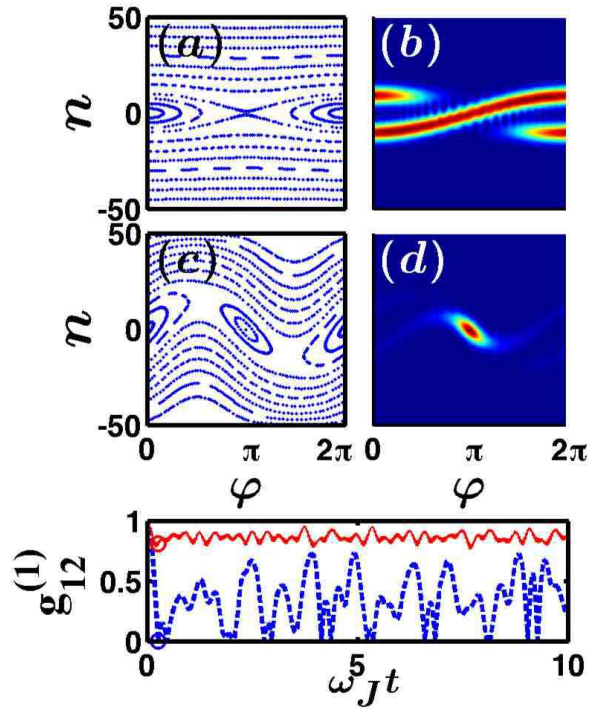}
\caption{(Color online) 
Quantum Kapitza pendulum with vertical driving,  
and ${|0,\pi\rangle}$ coherent preparation. 
The panels are arranged as in Fig.~1. 
The parameters are $u{=}100$, $\Omega{=}30$, and $N{=}100$.
The strength of the driving is ${q{=}0}$ (a,b, lower panel dashed blue) 
and ${q{=}3}$ (c,d, lower panel solid red). Circles denote the time at which 
the Husimi distribution in (c,d) is plotted. The classical stabilization of the 
inverted pendulum results in a protected 
single-particle coherence of the initial preparation.}
\label{Vertical}
\end{figure}
%%%%%%%%%%%%%%%%%%%%%%

%fig4
%%%%%%%%%%%%%%%%%%%%%%%
\begin{figure}[t]
\centering
\includegraphics[angle=0,width=0.50\textwidth]{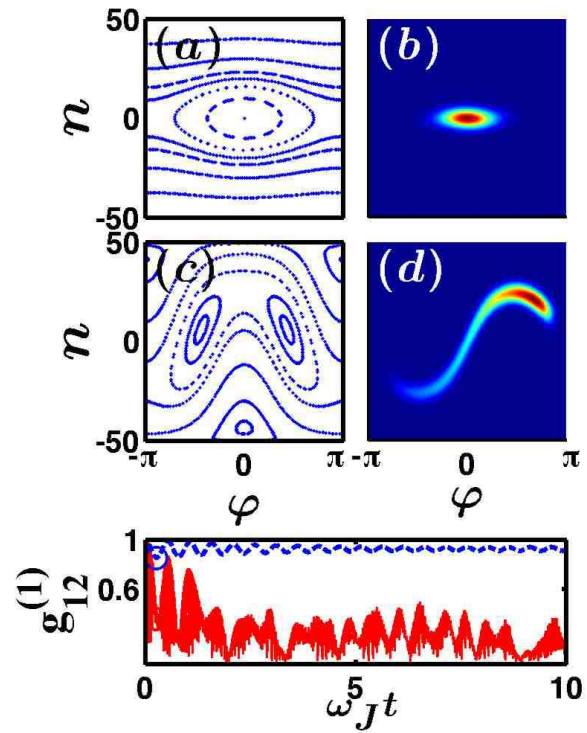}
\caption{(Color online) Quantum Kapitza pendulum with horizontal driving, 
and ${|0,0\rangle}$ coherent preparation. 
The panels are arranged as in Fig.~3, with the same parameters, except $u=30$.
One observes the splitting of the ground state.  
Due to this destabilization the fringe visibility of the preparation is destroyed.}
\label{Horizontal}
\end{figure}
%%%%%%%%%%%%%%%%%%%%%%

The predicted Kaptiza physics effects are confirmed numerically in Fig.~\ref{Vertical} and Fig.~\ref{Horizontal} for the vertical $V_v(t)$ and horizontal $V_h(t)$ drive, respectively. Comparison of the stroboscopic Poincare plots for the undriven (a) and driven (c) BHH, clearly shows the stabilization of the $|0,\pi\rangle$ coherent state by the vertical~$\Lx$ driving (Fig.~\ref{Vertical}), and the destabilization of the $|0,0\rangle$ preparation by the horizontal~$\Ly$ driving (Fig.~\ref{Horizontal}). These classical effect are mirrored in the evolution of the quantum Husimi function, thereby affecting the many-body fringe-visibility dynamics, leading to the  protection of coherence by $V_v(t)$ driving for $\varphi=\pi$ coherent preparation, and to its destruction by $V_h(t)$ for $\varphi=0$. 

%%%%%%%%%%%%%%%%%%%%%%%%%%%%%%%%%%%%%%%%%%%%%%%%%%%%%%%%%%%%%%%%%%%%%%%%%%%%%%
%%%%%%%%%%%%%%%%%%%%%%%%%%%%%%%%%%%%%%%%%%%%%%%%%%%%%%%%%%%%%%%%%%%%%%%%%%%%%%
% conclusions

To conclude, the driven BHH, currently attainable in a number of experimental setups, presents a wealth of nonlinear phase-dynamics effects. Strong, resonant driving fields result in large chaotic phase-space regions, opening the way for the generation of exotic highly correlated quantum states \cite{Ghose01}. The properties of such states, as well as their manifestation in interference experiments, and the more conventional tunneling effects between regular islands, are novel manifestations of semiclassical physics. For weak and fast off-resonant drive we have obtained the many-body equivalents of the Kapitza pendulum effects, with the relative-phase between the condensates acting as the pendulum angle. Such effects could be readily observed in interference experiments and utilized to protect fringe-visibility.
We note that noise-protected coherence was also studied in Ref.~\cite{Khodorkovsky08}, yet with a rather different quantum-Zeno underlying physics.

%The fringe-visibility in a many-realizations interference experiment is defined as $g^{(1)}_{12}\equiv2|\langle\eidag_1 \ei_2\rangle|/N=\sqrt{|\langle\Lx\rangle|^2+|\langle\Ly\rangle|^2}/\ell$ 

%%%%%%%%%%%%%%%%%%%%%%%%%%%%%%%%%%%%%%%%%%%%%%%%%%%%%%%%%%%%%%%%%%%%%%%%%%%%%%
%%%%%%%%%%%%%%%%%%%%%%%%%%%%%%%%%%%%%%%%%%%%%%%%%%%%%%%%%%%%%%%%%%%%%%%%%%%%%%

We thank Saar Rahav \cite{Rahav03} for usefull communication. This work was supported  by the Israel Science Foundation (Grant 582/07), by grant nos. 2006021, 2008141 from the United States-Israel Binational Science Foundation (BSF), and by the National Science Foundation through a grant for the Institute for Theoretical Atomic, Molecular, and Optical Physics at Harvard University and Smithsonian Astrophysical Observatory.

\vspace*{-4mm}

%%%%%%%%%%%%%%%%%%%%%%%%%%%%%%%%%%%%%%%%%%%%%%%%%%%%%%%%%%%%%%%%%%%%%%%%%%%%%%
%%%%%%%%%%%%%%%%%%%%%%%%%%%%%%%%%%%%%%%%%%%%%%%%%%%%%%%%%%%%%%%%%%%%%%%%%%%%%%

%%%%%%%%%%%%%%%%%%%%%%%%%%%%%%%%%%%%%%%%%%%%%%%%%%%%%%%%%%%%%%%%%%%%%%%%%%%%%%
%%%%%%%%%%%%%%%%%%%%%%%%%%%%%%%%%%%%%%%%%%%%%%%%%%%%%%%%%%%%%%%%%%%%%%%%%%%%%%

\end{document}